\documentclass[12pt, a4paper,twoside]{article}
\usepackage[super,sort&compress,comma]{natbib}
\usepackage[version=3]{mhchem}
\usepackage{times,mathptmx}
\usepackage{setspace}
\AtBeginDocument{\doublespacing}
\usepackage{sectsty}
\usepackage{balance}

\usepackage{graphicx} 
\usepackage{lastpage}
\usepackage[format=plain,justification=raggedright,singlelinecheck=false,font=small,labelfont=bf,labelsep=space]{caption}
\usepackage{fancyhdr}
\usepackage[version=3]{mhchem}

\author{%
B. J. Nagare\thanks{Department of physics,
University of Mumbai, Santacruz(E)
Mumbai-400 098,
India. Email:\texttt{bjnagare@gmail.com}}
\and
Darshan Habale\thanks{%
Department of physics,
University of Mumbai, Santacruz(E)
Mumbai-400 098,
India.}
\and
Sajeev Chacko\thanks{%
Inter-University Acceleration Centre,
Aruna Asaf Ali Marg, Near Vasant Kunj, New Delhi-110067,
India. Email: \texttt{sajeev.chacko@gmail.com}}
\and
Swapan Ghosh\thanks{%
Bhabha Atomic Research Center,
Department of Theoretical Chemistry,
BARC, Trombay,
Mumbai - 400085,
India.}
}
\title{Hydrogen Adsorption on Na/SWCNT Systems}
\begin{document}
\maketitle
\begin{abstract}
We investigate the hydrogen adsorption capacity of Na-coated carbon nanotubes (Na-SWCNTs) 
using first-principles electronic structure calculations at absolute temperature and pressure. 
A single Na atom is always found to occupy the hollow site of a hexagonal carbon ring in all 
the six different SWCNTs considered, with a nearly uniform Na-C bond length of~2.5 A. 
Semiconducting zigzag nanotubes, (8,0) and (5,0), show stronger binding energies for the Na atom
(-2.1~eV and -2.6~eV respectively), as compared to metallic SWCNTs with armchair and chiral 
geometries.  The single Na atom can further adsorb up to six hydrogen molecules with a relatively 
constant binding energy of ~-0.26~eV/H$_{2}$. Mulliken population analysis shows that positively 
charged Na atoms with 0.82$e$ charge transfer to nearest carbon atoms which polarizes the SWCNT
leading to local dipole moments. This charge-induced dipole interaction is responsible for the higher 
hydrogen uptake of Na-coated SWCNTs.  The transition state search shows that diffusion barrier of Na 
atom on the SWCNT between two adjoining C-C rings is ~0.35~eV.  We also investigate the clustering of 
Na atoms to find out the maximum weight percentage adsorption of H$_{2}$ molecules. At high Na 
coverage, we show that Na-coated SWCNTs can adsorb 9.2-11.28~wt~$\%$ hydrogen.  Our analysis shows 
that, although indeed Na-coated SWCNTs present potential material for the hydrogen storage, care 
should be taken to avoid Na atoms clustering on support material at elevated temperature and pressure,
to achieve higher hydrogen capacity.
\end{abstract}
\section{Introduction}
Recent advances in materials science and technology have witnessed a rapid expansion of research towards 
discoveries of novel materials for sustainable energy. In this cause, hydrogen has been identified as an 
alternative energy carrier because of its high abundance, lightweight, and environmental 
friendliness~.\cite{sch-01} Yet, there are still some important challenges to be addressed prior to possible 
actual spread of hydrogen-based infrastructure. One of the critical problems is lack of suitable hydrogen 
storage materials which can store H$_{2}$ with high volumetric and gravimetric density, and also desorbs 
H$_{2}$ through easy means. Furthermore, hydrogen adsorption and desorption should be operated under near 
ambient conditions in cost effective ways.~\cite{meng-07, xiong-08} Since the first experiment of hydrogen 
adsorption on carbon nanotubes by Dillon \textit{et al.}~\cite{nature-Dil97} and announcement of hydrogen 
storage target of 6$\%$ by weight for on-board automotive applications by the US Department of 
Energy~(DOE)\cite{heden-2000}, carbon nanotubes are considered as suitable candidates because of their light 
weight, high strength, high adsorption characteristics, and also large surface to volume ratio. However, experimental
findings showed that carbon nanotubes are not suitable for the H$_{2}$ adsorption due to weak van der Waal 
interaction between H$_{2}$ and carbon surface.\cite{nature-Dil97} These works have triggered extensive research 
on the functionalization of the carbon nanotubes by incorporating transition metals as well as alkali and 
alkaline earth metals.

The study of hydrogen storage using transition (Sc, Ti, and Ni), alkali (Li, K, Na) and alkaline earth 
(Mg, Ca) metals doped carbon nanotubes have been reported by many research 
groups~\cite{pavel-07,prl-Cir05,jpc-Liu09,science-Che99, jcp-Cab05,jacs-sun06,prl-yoon08}. A single 
transition-metal (Sc, Ti, and Ni) atom coated on C$_{60}$ or carbon nanotube surface can bind up to 
three to five H$_{2}$ molecules~\cite{prl-zhao05, prl-Cir05,apl-shin06}.  Thus, the dispersed C$_{60}$ and
nanotubes can adsorb up to 6.8-10~wt~$\%$ hydrogen with high metal coverage. Carbon nanomaterials decorated 
with alkali and alkaline-earth metal atoms can also be used for hydrogen storage.  Chen~\textit{et al.} have 
reported a hydrogen adsorption capacity of 20~wt\%\ and 14~wt\%\ in lithium and potassium doped carbon
nanotubes\cite{science-Che99} respectively.  Later on, it was found that observation of such high adsorption 
capacity was due to the presence of water impurities\cite{carbon-Yan00}. Cabria~\textit{et al.}~\cite{jcp-Cab05} 
have shown that the binding energy of hydrogen molecule on Li-doped carbon nanotubes is two times higher than 
that on pristine carbon nanotubes.  Recently, Liu and coworkers\cite{jpc-Liu09} found that up to eight Li metal 
atoms dispersed on single walled carbon nanotubes (SWCNTs) can adsorb 64~H$_{2}$ molecules with an average 
binding energy -0.17~eV/H$_{2}$ leading to 13.5~wt\%\ of storage capacity\cite{jpc-Liu09}. 
Ataca~\textit{et al.}~\cite{apl-Ataca08} have studied the dispersion of Li atoms on graphene surface and 
predicted the gravimetric density of 12.8~wt\%. The isolated clusters (Li$_{12}C_{60}$, Na$_{8}C_{60}$ and 
Ca$_{32}C_{60}$) where metal atoms are capped onto the pentagonal and hexagonal faces of C$_{60}$, can 
store H$_{2}$ up to 8.4-9.5~wt~$\%$\cite{jacs-sun06, nanolett-Gho08, prl-yoon08}. The charge transfer from
metal atoms to C$_{60}$ gives rise to the electric field surrounding the coated fullerene and enhances the 
H$_{2}$ adsorption. Recently, it is found that the sodium ions have been strongly absorbed onto a hydrophobic 
graphite surface via cation-$\pi$ interactions\cite{chinphys-shi11}.

We take the sodium as a coating material on SWCNTs since the sodium ion is one of the most popular ions 
in medical science, chemistry, physics and particularly in biological systems. It is sixth most abundant 
element in nature and most studied in the periodic table due to its simplicity towards the experimentation. 
On comparing the earlier experimental~\cite{langchak-95,assglee-04} and theoretical reports with the present 
study, it can be mentioned that the sodium-doped carbon nanotubes have many advantages and satisfy some of 
the important requirements for the hydrogen storage materials; namely,(a) strong interaction between the 
sodium ions and H$_{2}$ molecules, (b) remarkably enhanced molecular H$_{2}$ adsorption, and (c) light-weight 
material with higher capacity for the adsorption of H$_{2}$ molecules. The sodium-doped SWCNTs can act as a 
better hydrogen storage material at lower temperature with better hydrogen adsorption efficiency. Further, the 
adsorption can even be tuned by functionalizing with other suitably chosen organic functional groups~\cite{rscsri-12}. 
There are many reports on hydrogen storage using sodium for e.g., the metal complex of sodium (sodium borohydrates, 
sodium alanates\cite{chemmater-Zheng08, jamchemsoc-cornelis08}), sodium-doped fullerenes~\cite{nanolett-Gho08}, 
adsorption of sodium ions on hydrophobic graphite surfaces\cite{chinphys-shi11}, curvature-induced hydrogen 
adsorption in sodium-doped carbon nanomaterials\cite{chandra-08}. All above studies have been carried out at 
molecular levels. To the best of our knowledge, there is no report on sodium doped carbon nanotube at solid 
state. Despite of many reports for hydrogen storage on Na-doped carbon materials at molecular level,
there is no attempt to explain the basic understanding of the stability of sodium atoms on carbon surfaces and 
its high storage capacity towards H$_{2}$ molecules. So, the interesting issues we address are; (i). will the 
high storage capacity of Na-doped carbon nanotubes sustain at solid state? (ii). binding mechanism and stability 
of sodium atoms on carbon nanotube surfaces, (iii). evolution of the electronic properties of Na-SWCNT-H$_{2}$ systems,
(iv). adsorption mechanism of H$_{2}$ molecules on Na-SWCNTs, (v). effects of sodium clustering on the stability 
and adsorption of H$_{2}$ molecules, (vi). electrostatic interaction of sodium and H$_{2}$ molecules on carbon surface,
(vii). effects of chirality of carbon nanotubes on the stability and adsorption of sodium and H$_{2}$ molecules,
(viii). transition states search of composite systems (Na/SWCNT-H$_{2}$). To address the above issues, we carry 
out extensive search for high-capacity hydrogen storage material consisting of individual sodium atom doped carbon
nanotubes at zero temperature using density functional theory (DFT). We employ both local density approximation 
(LDA)~\cite{prb-Per81,prl-Cep80} and generalized gradient approximation (GGA)~\cite{prl-Per96} to estimate the
binding strength. We have also verified the LDA results with hybrid B3LYP functional~\cite{jcpbeck-88}.

Earlier studies\cite{jcp-cab08,jpcb-oka01} have found that the LDA based predictions of the physisorption 
energies of H$_{2}$ on the surface of graphite and carbon nanotubes were in good agreement with experiments.
Furthermore, quantum Monte Carlo calculations show that the adsorption energy of hydrogen in carbon materials 
typically lies between the calculated adsorption energy based on GGA and LDA\cite{prl-zhao05}.  A previous
\textit{ab initio} study using the M{\o}ller-Plesset second-order perturbation~(MP2) method\cite{jpcb-oka01} has
shown that GGA tends to underestimate the adsorption energy of H$_{2}$ on SWCNT whereas LDA gives reasonable 
adsorption energy as compared to the MP2 calculation\cite{jcp-cab08,prl-zhao05,jpcb-zho06}. The reliability 
of LDA can be ascribed to the following facts\cite{jcp-cab08}. Firstly, when the electron densities of H$_{2}$ 
and carbon surface overlap weakly, the nonlinearity of the exchange-correlation energy density functional produces 
an attractive interaction even in the absence of electron density redistribution. Also the overestimated binding 
energy by LDA\cite{prb-lug07, prb-lee08} may compensate for the insufficient account of van der Waals 
interactions.\cite{jcp-cab08} In contrast, DFT calculation using a uniform GGA produced a purely repulsive
interaction.  Calculation based on the GGA-PW91 functional has predicted a repulsive interaction between H$_{2}$ 
and a graphene layer and also between H$_{2}$ and a (6,6) carbon nanotube\cite{prb-tad01}, which contradicts the
experimental findings\cite{jmat-sah08}. It has also been noted that LDA calculations well reproduce the empirical 
interaction potentials between graphitic layers and also in the other graphitic systems for distances near
to the equilibrium separation although the LDA is not able to reproduce the long-range dispersion 
interaction\cite{prb-gir02}.

Accordingly, in the present work, the conventional DFT calculations are directed toward obtaining structural 
stability and electronic structure properties of  Na-SWCNT-H$_{2}$ systems at absolute temperature and pressure. 
The transition states of the reaction are also studied in order to determine the energy barrier of the reaction 
and the reaction rate. Semiconducting zigzag nanotubes show stronger binding energies for the Na atom, as
compared to metallic SWCNTs with armchair and chiral geometries. We show that the charge-induced dipole interaction 
is responsible for the higher hydrogen uptake of Na-coated SWCNTs. We find that the storage capacity of hydrogen 
depends strongly on the clustering of Na atoms on SWCNTs.
\section{Computational Details}
All the energy calculations and geometry optimizations were performed at absolute temperature by using density 
functional theory as implemented in the CASTEP code~\cite{zfk-Cla05}. We use the local density 
approximation~(LDA) with Perdew-Zunger parameterization~\cite{prb-Per81} of Ceperley-Alder data~\cite{prl-Cep80} 
as well as generalized gradient approximation~(GGA) by Perdew-Burke-Ernzerhof~(PBE)~\cite{prl-Per96}. Ultrasoft
pseudopotential with a plane wave basis cutoff of 380~eV was used. To ensure the calculated results being 
mutually comparable, the identical conditions are employed for the isolated H$_{2}$ molecules, Na atom,
the carbon nanotube, and also the adsorbed carbon nanotube system. The Monkhorst-Pack scheme with 1x1x7 special 
K-points is used for Brillouin zone sampling for all the systems. The geometries were considered converged 
when the change in energy was of the order of 10$^{-5}$~Hartrees and the absolute maximum force was 
0.002~Hartree/\AA.
\begin{figure}
\center
\includegraphics[scale=0.2]{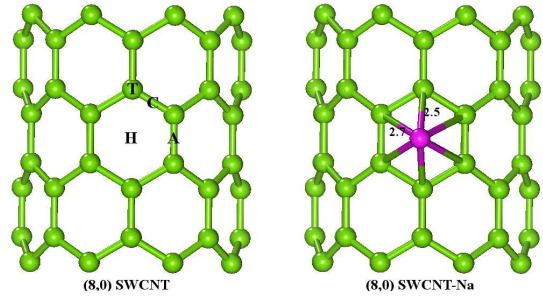}
\caption{Four possible Na adsorption sites on~(8,0) SWCNT. Symbols A, C, T, and H
denote the bridge site over an axial bond, the bridge over a zigzag C-C bond,
top of the carbon atom, and the hollow site of the hexagonal C-ring.
The light green and pink spheres represent carbon and sodium atoms.}
\label{fig-cnt-site}
\end{figure}

We have selected several carbon nanotubes with different chiral indices as the model of carbon substrate.
All simulations were carried out in periodicity along Z-axis. The dimensions of simulation cells ranged 
from 16$\times$16$\times$8.5 for smallest system to 27.4$\times$28$\times$7.38~\AA$^{3}$
for the largest systems respectively. The simulation cell were taken so that the empty space between 
the nanotubes and the X and Y directions were $\sim$~7.5~\AA. Further increase in this vacuum space 
leads to negligible change in the total energy of the systems and forces on the atoms. For most part of 
simulations, we have selected (8,0) SWCNT because of its feasibility for synthesis.
\begin{figure}
\center
\includegraphics[scale=0.15]{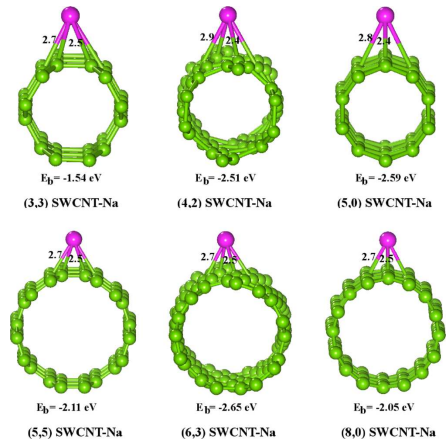}
\caption{Optimized geometries of Na adsorbed at hollow sites on (3,3), (4,2),
(5,0), (5,5) (6,3), and (8,0) SWCNTs. E$_{b}$ represents the binding energy of Na.
The light green and pink spheres represent carbon and Na atoms, respectively.}
\label{fig-cnt-na}
\end{figure}

A storage model composed of Na-doped (8,0) SWCNT is built by considering four possible Na adsorption
sites. These four possible sites are the top of the carbon atom (T), the hollow site of the hexagon
C-ring (H), the bridge site over an axial bond (A), and the bridge over a zigzag C-C bond (C) as shown 
in figure 1.  After full structural optimization, the Na atom always locate on hollow site (H) shown 
in figure 2, regardless of the initial location. The binding energy of the Na atom is given by,
\begin{equation}
BE_{Na} = E_{Na-SWCNT}-(E_{SWCNT} + E_{Na})
\end{equation}
where E$_{Na-SWCNT}$, E$_{SWCNT}$ and E$_{Na}$ are the total energies of Na-doped SWCNT, SWCNT and 
isolated Na atom respectively. Next, we used lowest energy structure of Na-SWCNT for H$_{2}$ adsorption.
The binding energy per H$_{2}$ molecule is given by,
\begin{equation}
BE_{H_2} = [E_{(H_2)_n Na-SWCNT} - E_{Na-SWCNT}-nE_{H_2}]/n
\end{equation}
where n indicates the number of H$_{2}$ molecules, E$_{(H_2)_n -Na-SWCNT}$, E$_{Na-SWCNT}$ and E$_{H_2}$ are the
total energy of (H$_{2})_n $-Na-SWCNT complex, Na-doped SWCNT and H$_{2}$ molecule respectively. The adsorption 
mechanisms are analyzed in terms of the Mulliken population analysis, electrostatic potential and partial
density of states (PDOS) which can provide a clear and definitive description for charge distribution. We 
have verified the accuracy of our calculations by calculating the density of states (DOS) of (8,0)SWCNT and 
the results agree well with reported results of Li \textit{et al}.~\cite{jcp-li03}
\section{Results and Discussions}
\label{sec-results}
First, we consider the adsorption energetics of a single Na atom on the (8,0) SWCNT as shown in figure 2. 
We have performed structural optimizations starting with four possible metal adsorption sites (A, C, H, and 
T) on carbon nanotubes. A single Na atom is found to always occupy the hollow site of a hexagonal carbon 
ring in all the six different CNTs considered, with Na-C bond length of 2.4-2.9 \AA. 
\begin{figure}
\center
\includegraphics[scale=0.8]{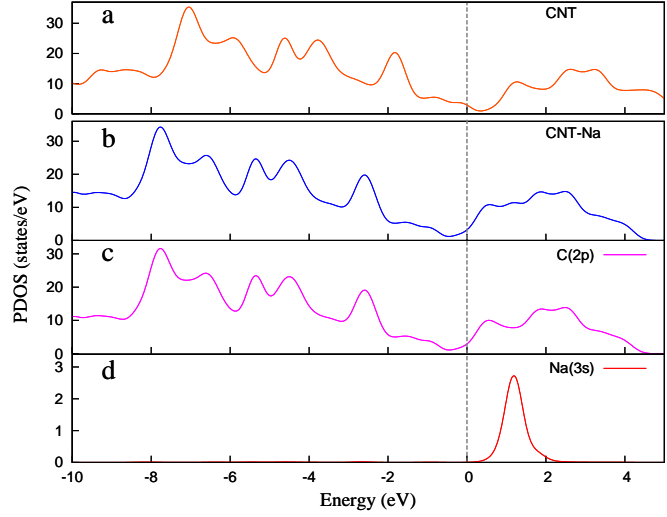}
\caption{The density of states of (8,0) SWCNT and SWCNT-Na are shown. The partial density of states of 
carbon~(2$p$) and sodium~(3$s$) atoms of (8,0)~SWCNT-Na system are also shown. The Fermi level is set to 
zero and indicated by the dotted line.}
\label{fig-na-pdos}
\end{figure}
When a Na atom is adsorbed on SWCNT, there are two nearest and four next-nearest neighbors. It was believed 
that the stability of adsorption was enhanced by the curvature of the carbon nanotube surface\cite{chandra-08}.
Here, we have considered six SWCNTs [(3,3), (4,2), (5,0), (5,5), (6,3) and (8,0)]. It is observed that 
E$_{b}$ for Na atoms on (5,0) and (8,0) tubes are -2.59~eV and -2.05~eV, while E$_{b}$ for Na atoms on the 
(3,3), (4,2), (5,5) and (6,3) are -1.54~eV, -2.51~eV, -2.11~eV and -2.65~eV respectively. The binding energy 
difference comes from the configuration of the surface. It is observed that Na atoms on semiconducting SWCNTs
[(4,2), (5,0),(6,3)and (8,0)] are more strongly adsorbed by $\sim$0.4~eV as compared with metallic 
SWCNTs [(3,3), (5,5)] due to strong electrostatic interaction between Na and SWCNTs.
\begin{figure}
\center
\includegraphics[scale=0.2]{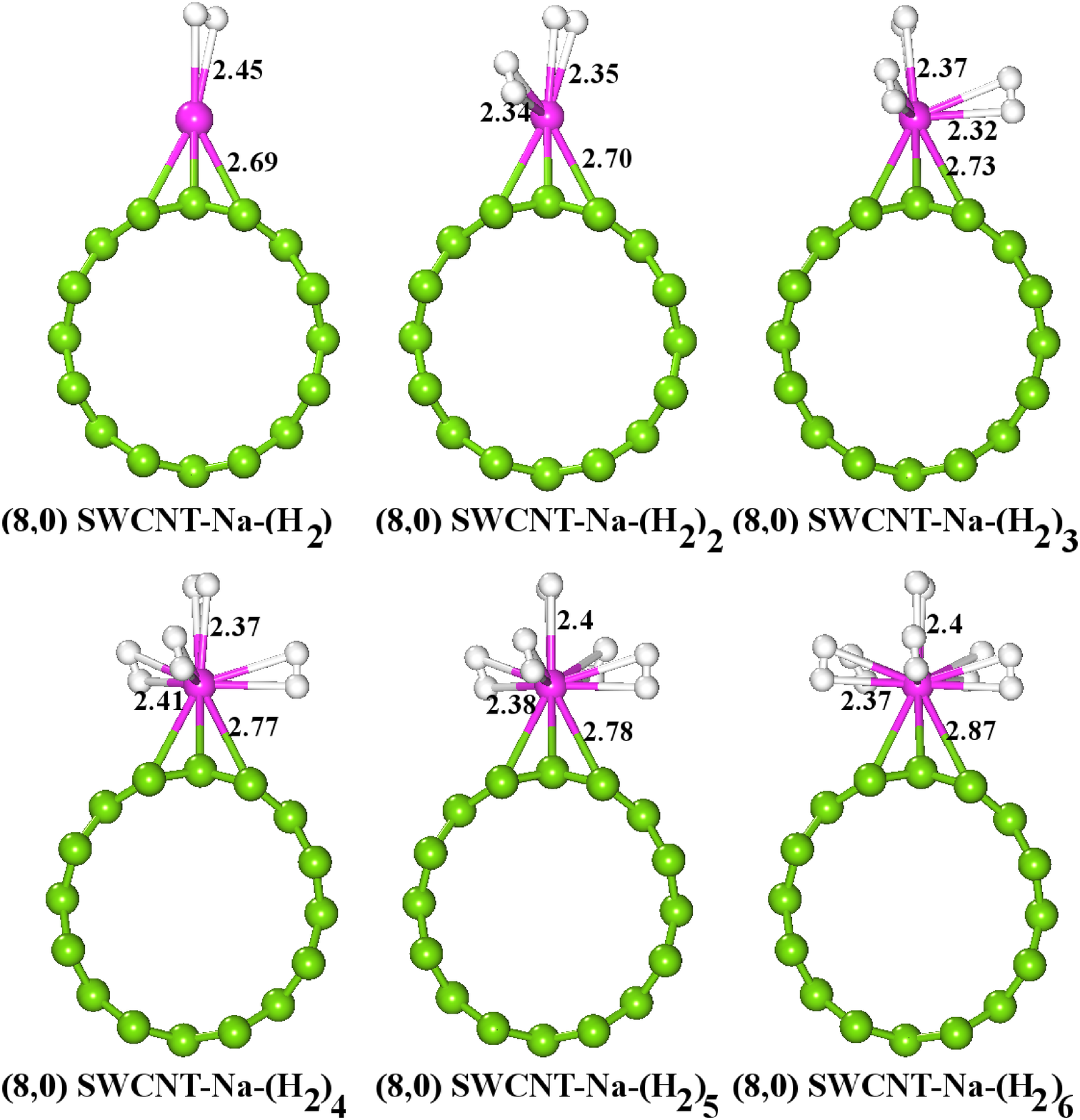}
\caption{The coverage of the H$_{2}$ molecules on (8,0) SWCNT-Na.
The light green and pink spheres have the same meaning Figure 2,
and the white one shows H atom.}
\label{fig-na-h2}
\end{figure}

Now, we address the underlying binding mechanism between Na atom and (8,0) SWCNT. A single Na atom prefers to adsorb
strongly at the H site, as shown in figure 2. The Na-C distances are 2.46~\AA~ and 2.70~\AA~for nearest and next-nearest
neighbor carbon atoms with the binding strength of E$_{b}$=-2.05~eV. The Mulliken charge analysis shows that Na 
carries a 0.82$e$ positive charge, indicating that the Na atom is ionized. The two nearest and four next-nearest 
carbon atoms are negatively charged with 0.18$e$ and 0.035$e$ respectively, whereas remaining charges are 
distributed over the whole SWCNT. The partial density of states (PDOS) in figure 3 shows a lowering of carbon 
eigenvalues by 0.77~eV for all the carbon 2$p$ orbitals. The charge around Na and the PDOS implies a simple 
electrostatic attraction as a binding mechanism between charged Na atom and SWCNT.
\begin{figure}
\center
\includegraphics[scale=0.8]{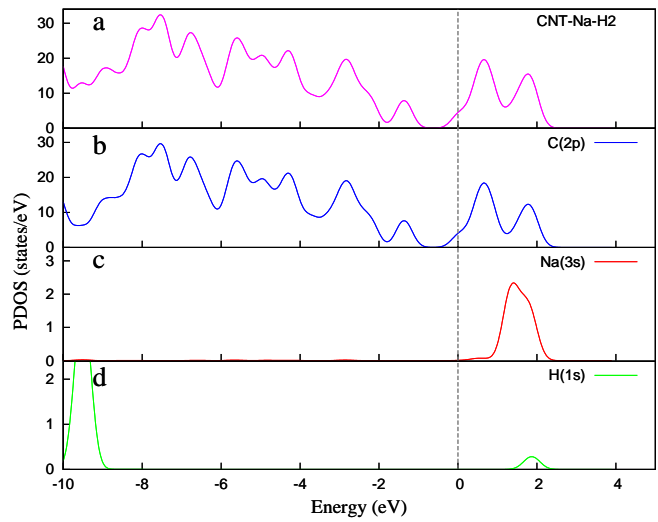}
\caption{The density of states and the partial density of states for C~(2$p$), Na(3s) and H$_{2}$(1s) 
of (8,0) SWCNT-Na-H$_{2}$ are shown. The Fermi level is set to zero and indicated by the dotted line.}
\label{fig-na-h2-pdos}
\end{figure}

Next, we investigate the interaction between Na-SWCNT with H$_{2}$ molecules. A single H$_{2}$ molecule 
sits atop the Na atom in (8,0) SWCNT-Na-H$_{2}$. The optimized structure is shown in figure 4. The molecular 
hydrogen is found to interact with Na atoms as a dihydrogen molecule with a binding energy -0.18~eV by LDA 
and Na-H bond length 2.44~\AA. Mulliken population analysis shows  charges of +0.76$e$ on the sodium and
+0.03$e$ on the hydrogen. From the total density of states (TDOS) and PDOS given in figure 5, the hydrogen 
bonding orbitals at $\sim$-9.5~eV far below the entire carbon 2$p$ band, hence there is no further lowering 
of carbon 2$p$ eigenvalues as compared to the SWCNT-Na system.
\begin{table}
\center
\small
\begin{center}
\caption{The bond length, charges on Na and binding Energy (eV)
of Na and H$_{2}$ on (8,0) SWCNT-Na}
\begin{tabular}{|c|c|c|c|c|c|c|c|c|}
\hline
No. of H$_2$ &  \multicolumn{2}{c|}{C-Na bond length (\AA)} &  \multicolumn{2}{c|}{Na-H bond length (\AA)}
& Charge on Na (e)  &       \multicolumn{3}{c|}{B.E. (eV)}       \\      \hline
molecules  & Minimum       & Maximum      & Minimum      &       Maximum &       &       LDA     &
GGA     &       B3LYP   \\      \hline
0     &       2.50   &       2.70   &       -       &       -       &       0.82
&       -2.05   &       -1.32   &       -1.38   \\
1    &       2.44   &       2.69   &       2.44   &       2.44   &
0.76   &       -0.18   &       -0.16   &       -0.23   \\
2    &       2.46   &       2.70   &       2.31   &       2.34   &
0.68  &       -0.27   &       -0.16   &       -0.21   \\
3    &       2.48   &       2.72   &       2.33   &       2.35   &
0.60   &       -0.26   &       -0.14   &       -0.21   \\
4    &       2.48   &       2.76   &       2.34   &       2.41   &
0.55   &       -0.26   &       -0.13   &       -0.21   \\
5    &       2.49   &       2.78   &       2.34   &       2.46   &
0.46   &       -0.26   &       -0.12   &       -0.21   \\
6    &       2.59   &       2.87   &       2.34   &       2.52   &
0.36   &       -0.26   &       -0.12   &       -0.20   \\       \hline
\end{tabular}
\end{center}
\end{table}

We also compute the electrostatic potential of SWCNT-Na-H$_{2}$ system to understand the binding mechanism 
of H$_{2}$. The substantial charge redistribution upon Na coating leads to high a electric field near the Na 
atoms. We measure the electric fields, induced by the charge redistribution along the outward radial direction of
the complex. Figure 6 shows the radial component of the electric field from the center of the SWCNT to the
hollow site of SWCNT, Na atom and H$_{2}$ molecule. For the SWCNT, the electric field strength is 
8.3x10$^{10}$ V/m, at the hollow site; with Na adsorbed, E(Na site) is 29.2x10$^{10}$ V/m; with single 
H$_{2}$ coverage atop Na, the E~(Na site) drops to 24x10$^{10}$~V/m and at the center of H$_{2}$ molecule 
E(H$_{2}$) is 8.6x10$^{10}$ V/m. Therefore, we assume that the electric field produced by SWCNT-Na polarizes 
the H$_{2}$ molecules and the induced dipole-dipole as well as the charge-dipole interactions, consequently, 
leads to binding of H$_{2}$ around SWCNT-Na. The PDOS of SWCNT-Na-H$_{2}$ shows that there is no bonding
orbital between H$_{2}$ and Na, which also supports this argument.
\begin{figure}
\center
\includegraphics[scale=0.9]{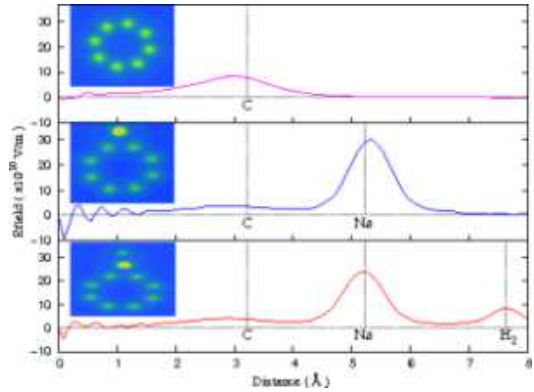}
\caption{The radial component of the electric field of (8,0) SWCNT, SWCNT-Na and SWCNT-Na-H$_{2}$ systems.
The dotted vertical lines indicate the position of Carbon ring, Na atom and H$_{2}$ molecules. The insets 
shows the isosurfaces of the electrostatic potentials of SWCNT, SWCNT-Na and SWCNT-Na-H$_{2}$.}
\label{fig-na-h2-esp}
\end{figure}

The optimized geometries of higher coverage of H$_{2}$ molecules on the (8,0) SWCNT-Na are shown in figure 4. 
The hydrogen binding energies (E$_{b}$) are summarized in Table 1, which show the GGA, LDA and B3LYP results 
for both Na atom and H$_{2}$ molecules. A single Na atom is found to adsorb up to six~H$_{2}$~molecules with 
a relatively constant binding energy of~-0.26~eV/H$_{2}$. It may be interesting to note that the maximum number 
of H$_{2}$ adsorbed at the Na site is very much similar to the case of Na-doped fullerene. Electrons in a 
carbon nanotube undergo major changes in their charge distribution upon Na doping. The transferred charges 
are spread over the hexagon formed out of carbon atoms nearest to the Na atom. The significant charge 
redistribution results in an electric field strong enough to attract up to six H$_{2}$ molecules with a 
binding energy of $\sim$-0.26~eV per H$_{2}$ with LDA and -0.14 with GGA functionals. It is known that GGA 
generally underestimates the binding strengths whereas LDA overestimates them. To verify our results, we have 
carried out calculations using the hybrid functional, B3LYP, since it has been successfully used to study the 
hydrogen storage in alkali metal doped C$_{60}$. We find that the LDA energies are comparable to those obtained 
by B3LYP, which are almost twice of the same computed by GGA.~\cite{nanolett-Gho08} The optimized structures 
show that none of the relevant physical quantities, such as average H$_{2}$ binding energy, the Na-H bondlength 
and the H-H distance undergoes significant changes as the number of adsorbed H$_{2}$ molecules increases.
However, there is monotonous decrease in the net charge on the sodium atom up to about 0.36$e$ for six H$_{2}$ 
molecules. We have observed that all the six H$_{2}$ molecules are absorbed at the optimum distance as shown 
in Table 1. The optimum distance is an important parameter for estimating the hydrogen storage capacity of any 
material. It determines a finite-scale region for storing more H$_{2}$ molecules around the doped atoms. So a 
large distance is advantageous to reduce the repulsion between the adsorbed molecules. It may be noted that the 
hydrogen molecules are found to interact with ionized Na atom as a dihydrogen molecule, forming a T-shaped complex,
and thus retaining their molecular identity irrespective of the type of SWCNTs.
\begin{table}
\small
\center
\begin{center}
\caption{Binding Energies (in~eV) of Na and H$_{2}$ molecules on (3,3), (4,2), (5,0), (5,5), (6,3) and (8,0) 
SWCNT-Na systems}
\begin {tabular}{|c|c|c|c|c|c|c|c|c|c|c|}
\hline
No. of H$_2$ &
\multicolumn{2}{c|}{(3,3)SWCNT-Na}    &       \multicolumn{2}{c|}{(4,2)SWCNT-Na}      &
\multicolumn{2} {c|}{(5,0)SWCNT-Na}   &       \multicolumn{2} {c|}{(5,5)SWCNT-Na}     &
\multicolumn{2}{c|}{(6,3)SWCNT-Na}    \\      \hline
molecules&       LDA     &       GGA     &       LDA     &       GGA     &       LDA  &       GGA
        &  LDA  &       GGA     &       LDA     &       GGA     \\      \hline
0       &       -1.54   &       -1.04   &       -2.51   &       -1.71   &       -2.59 &
-1.69   &       -2.11   &       -1.66   &       -2.65   &       -1.83   \\
1       &       -0.22   &       -0.10   &       -0.24   &       -0.12   &       -0.20 &
-0.11   &       -0.30   &       -0.13   &       -0.25   &       -0.13   \\
2       &       -0.20   &       -0.10   &       -0.24   &       -0.13   &       -0.25 &
-0.12   &       -0.28   &       -0.14   &       -0.26   &       -0.13   \\
3       &       -0.20   &       -0.15   &       -0.25   &       -0.10   &       -0.24 &
-0.11   &       -0.28   &       -0.11   &       -0.26   & -0.11 \\
4       &       -0.21   &       -0.14   &       -0.25   &       -0.11   &       -0.25 &
-0.13   &       -0.28   &       -0.10   &       -0.27   &       -0.10   \\
5       &       -0.23   &       -0.11   &       -0.25   &       -0.10   &       -0.25 &
-0.10   &       -0.26   &       -0.10   &       -0.27   &       -0.11   \\
6       &       -0.22   &       -0.10   &       -0.25   &       -0.09   &       -0.23 &
-0.08   &       -0.25   &       -0.09   &       -0.26   & -0.10 \\      \hline
\end{tabular}
\end{center}
\end{table}

We have investigated the effects of nanotube chiral index (metallic vs semiconducting) of SWCNTs on H$_{2}$
adsorption. Accordingly, we consider (8,0) and (5,5) SWCNT which are of nearly same diameter. It is observed 
that the calculated binding energies per H$_{2}$ molecule on Na-doped (8,0) SWCNT are very close to that for 
(5,5) SWCNT for all $n$=1,...,6 coverage of H$_{2}$. This indicates that the binding of H$_{2}$ to the 
SWCNT-Na complex has little to do with the chirality of nanotubes. It is, however, important to know if the 
results reported above for (8,0) and (5,5)~SWCNT hold for other nanotubes as well and if they depend on the 
chirality. Here, besides (8,0) and (5,5), we have studied four other different SWCNT-Na-(H$_{2}$)$_n$, $i.e.$, 
corresponding to (3,3), (4,2), (5,0), and (6,3)~SWCNT. The optimized geometries of SWCNT-Na-(H$_{2}$)$_n$ 
show some regularity. We observe that the H$_{2}$ molecules get adsorbed on and around the Na atom site.
Thus, one H$_{2}$ molecule is always found to sit atop the Na atom, while the remaining five H$_{2}$ molecules 
sit around the Na atom and the H-H bonds are oriented tangential to SWCNT surface which form T-shape complex 
(see figure 4). This holds true for all the SWCNT-Na-(H$_2$)$_n$ systems. Table 2 shows the binding energy 
data for Na and H$_{2}$ molecules. These results show that although the binding energy of Na is different on 
metallic and semiconducting nanotubes, it doesn't affect the binding and the uptake capacity of H$_{2}$ molecules. 
In conclusion, the phenomenon that a single Na atom adsorbed on hollow site of SWCNT can bind up to six
molecular hydrogen is  very general and holds for all the geometries of nanotubes considered here and for 
other carbon-based nanomaterials such as C$_{60}$ as reported earlier.~\cite{nanolett-Gho08}  In this
context, it may however be noted that different types of fullerenes doped with alkali metal atoms do not 
necessarily adsorb the same number of H$_{2}$ molecules, where the number varies from 3 to 6. In theses 
cases, curvature seems to play a more significant role.
\begin{figure}
\center
\includegraphics[scale=0.2]{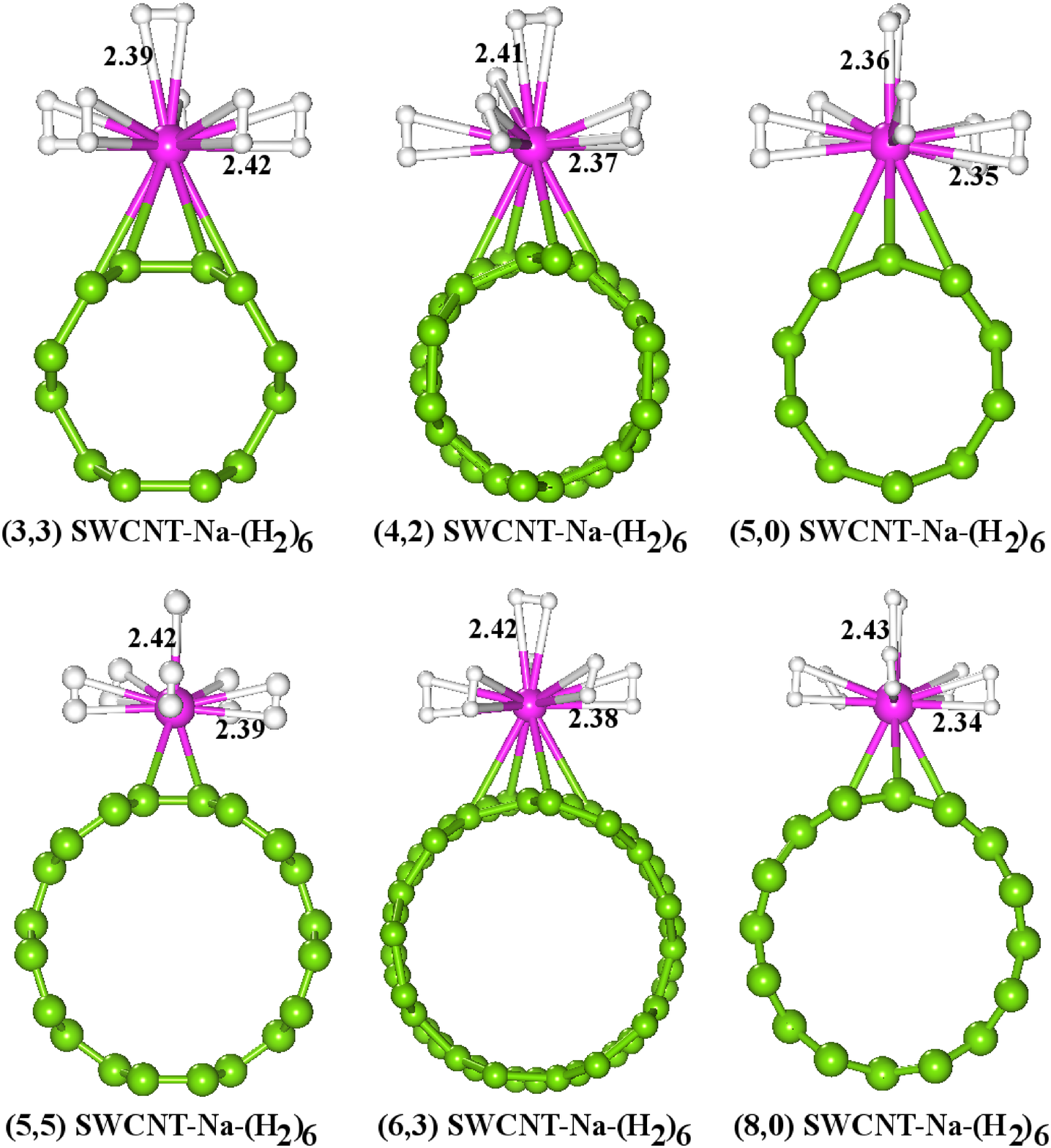}
\caption{The coverage of the H$_{2}$ molecules on (3,3), (4,2), (5,0), (5,5), (6,3) and (8,0) SWCNT-Na systems.}
\label{fig-cnt-all-na-h2}
\end{figure}

Now, we discuss the stable Na adsorbates with maximum coverage on SWCNT. If the sodium atoms are doped at 
all available hexagonal sites, it is expected that 144H$_{2}$ molecules can be adsorbed by the (8,0) 
SWCNT-24Na system. However, it is important to note that the adsorption of the maximum number of H$_{2}$ 
molecules at one cationic site is only possible when the cationic sites are separated by $\sim$5.0~\AA, 
at which the system can experience minimum electrostatic repulsion and steric hindrance. Accordingly,
we have identified such sites. In figure 8, we show six cases where Na covers 1/3 area of the hollow 
sites of (3,3), (4,2), (5,0), (5,5), (8,0) SWCNTs and 1/4 hollow sites of (6,3) SWCNT. The figure also 
shows the binding energy per Na and the Na-Na distances. The calculated binding energy values show that 
there is strong binding between the metal atoms and the SWCNT systems. We first discuss the (8,0)~SWCNT for
maximum coverage of Na. When Na atoms are adsorbed on the surfaces of SWCNT, there is significant charge 
transfer from the sodium atoms to the carbon atoms. As the coverage become higher (maximum eight Na atoms), 
the electrostatic repulsion increases due to shortening of the Na-Na distances which makes the system unstable.
(Recall that there is a charge transfer from sodium atoms to SWCNT making the sodium positively charged.)
For stable Na adsorbates on carbon nanotubes, the Na-Na distance should be larger than that in bulk Na 
which is 3.12~\AA.
\begin{figure}
\center
\includegraphics[scale=0.9]{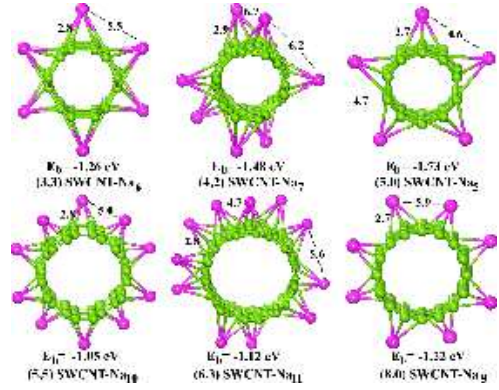}
\caption{The coverage of Na atoms on (3,3), (4,2), (5,0), (5,5), (6,3) and (8,0) SWCNT-Na systems.}
\end{figure}
For the stable Na adsorbates on carbon nanotubes, the Na-Na distance is 5.82~\AA~ with a binding energy 
of -1.22~eV per Na (In case of single Na, the binding energy is -2.05~eV). The decrease in binding energy 
can be understood from the net charge on sodium atoms. It is found that as the number of Na atoms increases, 
the charge on each Na atoms decreases to $\sim$0.48$e$ leading to decrease in binding energy per Na atom.
The calculated binding energy per Na atom is however larger than the cohesive energy of -1.13~eV of bulk Na.
This shows that clustering of Na atoms is unlikely to occur on SWCNT surface. It may be noted that when the 
Na-Na distance is less than 5.1~\AA, the net hydrogen molecule adsorption on Na-SWCNT decreases which is 
consistent with previous reported result\cite{nanolett-Gho08}.
\begin{figure}
\center
\includegraphics[scale=0.22]{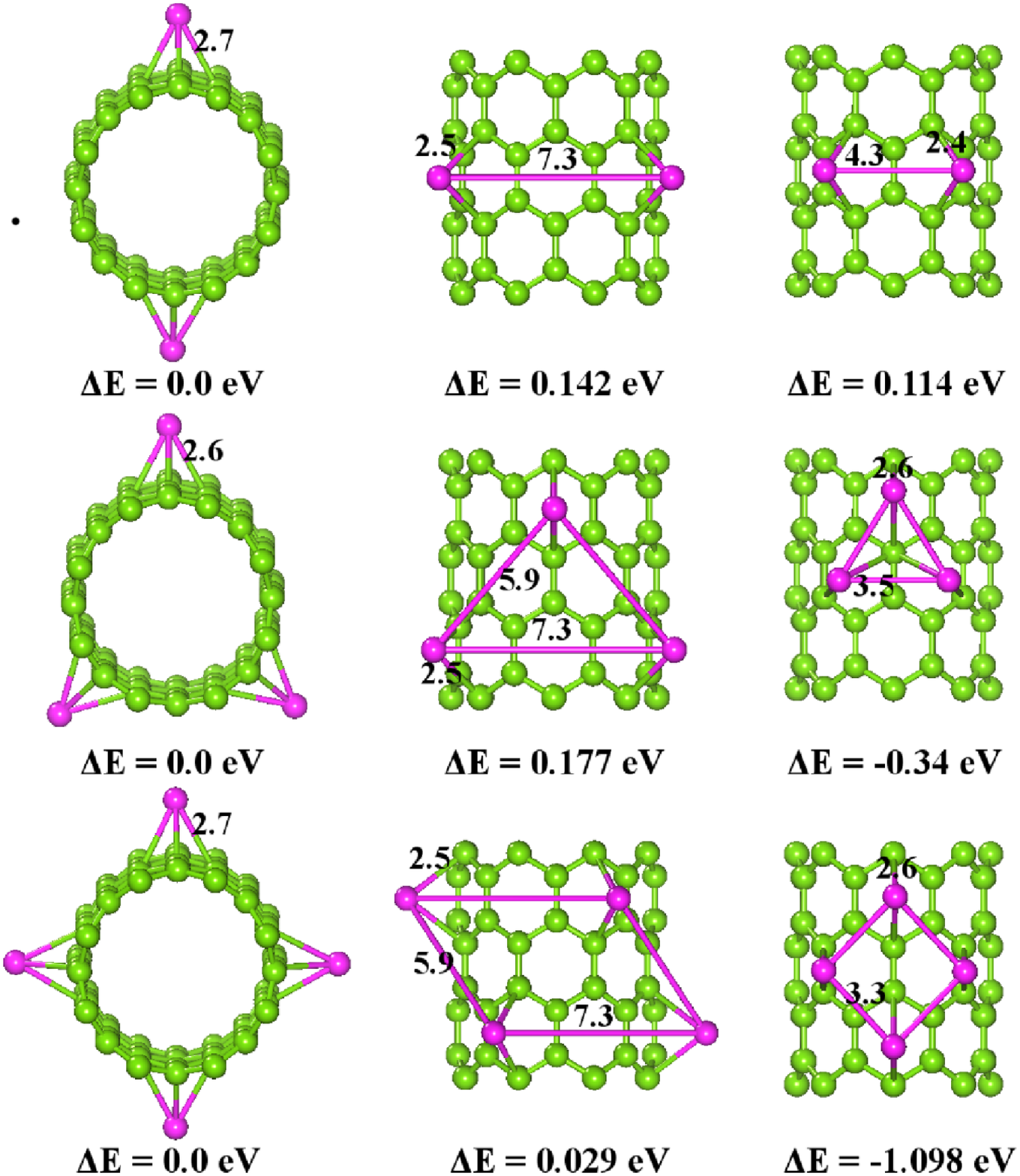}\\
\includegraphics[scale=0.20]{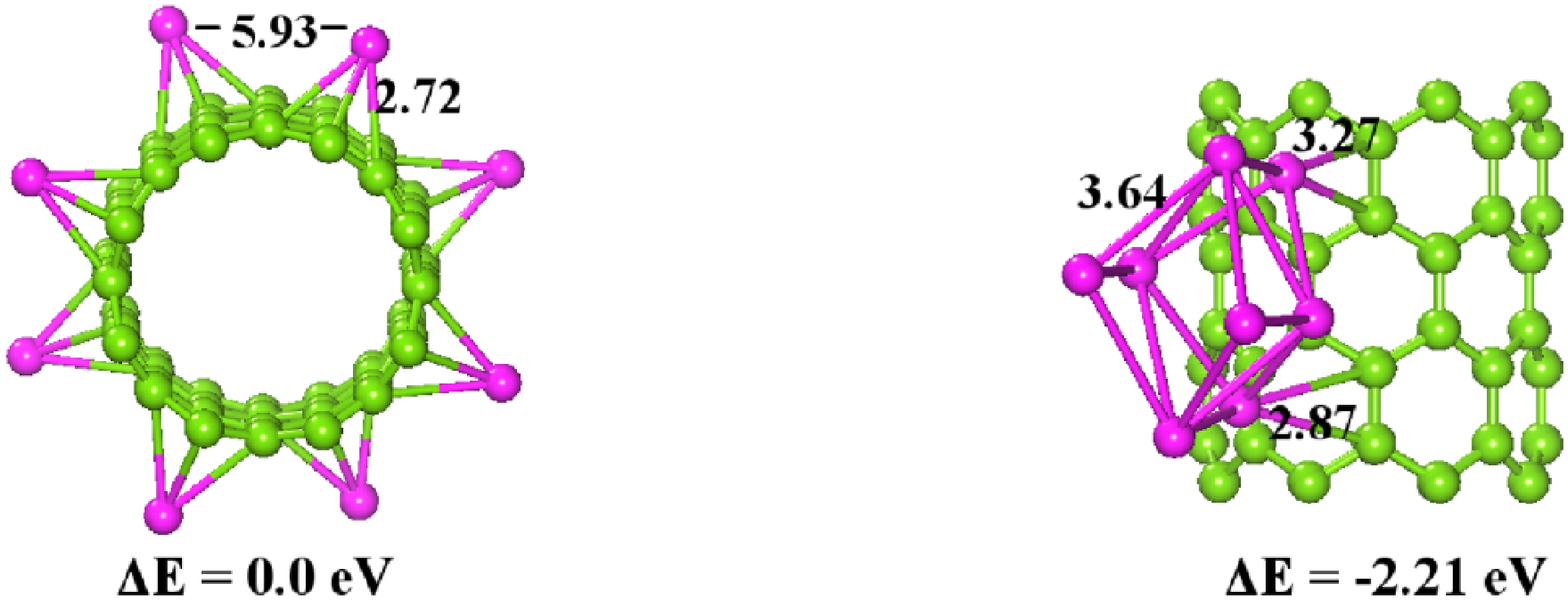}
\caption{Optimized geometries of the two, three, four and eight Na atoms isolated and aggregated on a 
pristine (8,0)SWCNT. ${\Delta}$E represent the energy difference with reference to ground state.}
\end{figure}

We now address the important of issue of Na clustering which directly determines the reversibility of hydrogen 
adsorption and desorption. To address this, we have considered the clusters of 2-8 Na atoms on the surface 
of (8,0) SWCNT as shown in figure 9. To check whether Na atoms form a cluster on the (8,0) SWCNT surface,
we have carried out two separate calculations. First, we placed two Na atoms at the hollow sites of next 
nearest neighbor carbon hexagon such that their separation distance is 7.3~\AA ~(column two of figure 9). 
In the second case, the Na atoms are on two adjacent carbon hexagon with a Na-Na separation distance of 
4.3~\AA~(column 3 of figure 9) for $n$=2 and decreasing to 3.3 \AA~as the cluster size grows to $n$=8.
The total energy difference ($\Delta$E) relative to isolated Na coverage (column 1 of figure 9) shows that 
for n=2, the coating structures are more stable than those involving clustering, and are 0.14 and 0.11~eV 
lower in energy. The Na clustering geometries are however uniformly more stable than the isolated Na systems, 
for n~$\ge$~2 and the stability increases with $n$, from 0.34~eV for $n$=3 to 2.21~eV for n~=~8; whereas the 
Na coating/clustering structures of column 2 of figure 9 are less preferred to the isolated Na geometries. The 
binding energies, 0.27~eV/H$_{2}$ for Na$_{2}$ and 0.22~eV/H$_{2}$ for Na$_{4}$ complexes hardly changes from 
the saturation value $\sim$0.26~eV/H$_{2}$ for 6H$_{2}$ complexes. The above analysis shows that for the 
clustering of Na atoms to occur, the Na-Na distance must be $\ge$~ 5.9~\AA. It is further to be noted
that our calculated binding energy for Na (for Na$_{8}$) on (8,0) SWCNT (-1.22~eV) is larger than the 
cohesive energy of bulk Na, namely, -1.13~eV. Since the energy gained from the cohesion is lower than that 
from binding to SWCNT, clustering of Na atoms is unlikely to occur on SWCNT.
\begin{figure}
\center
\includegraphics[scale=0.24]{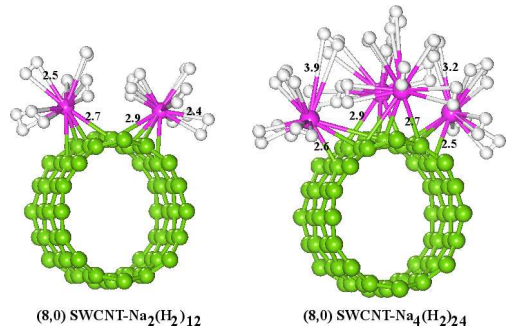}
\caption{Optimized geometries of the twelve and twenty four H$_{2}$ molecules on the two and four aggregated 
Na atoms on (8,0)SWCNT.}
\end{figure}

Now we discuss in brief the effect of Na clustering on the adsorption of the H$_{2}$ molecules. In Figure 
10, we show the optimized geometries for twelve and twenty four H$_{2}$ molecules on the aggregated Na$_{2}$ 
and Na$_{4}$ complexes. The optimized C-Na separation ranges from 2.5~\AA~to 3.0~\AA~. The clustering of Na has
thus little effect on the binding energy per H$_{2}$ molecules, and the uptake capacity of H$_{2}$ adsorption.
\begin{figure}
\center
\includegraphics[scale=0.8]{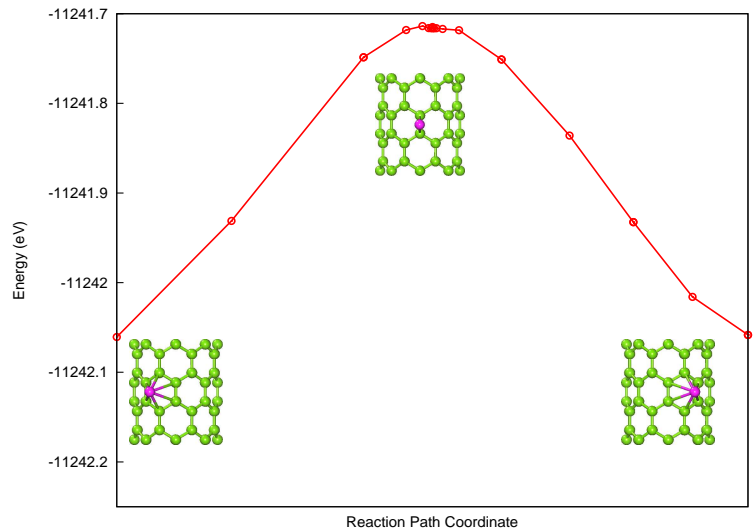}
\caption{Energy Vs reaction path for successive transformation of Na atom from one C-C ring to next nearest 
neighbor over a (8,0) SWCNT. Only one transition state is found during the reaction.}
\end{figure}

To use Na-doped SWCNT as a hydrogen storage materials, we have studied the transition state search using 
nudged elastic band method (NEB) to calculate the diffusion barrier which directly determines the adsorption 
and desorption mechanism of metal and hydrogen molecules. As the interaction between H$_{2}$ molecules and 
Na is physisorption in nature, we consider the transition state search for Na atom only. We calculate the 
diffusion barrier of Na atom between the two nearest neighbor C-C rings of the SWCNT. Through optimized search of the
transition state, only one transition state is found which is shown in figure~11. This transition state has 
the Na atom on the top of the bridge site of the two C-C rings. The diffusion barrier (0.34~eV) being larger than
0.26~eV/H$_{2}$, which is the binding energy, implies that during the desorption of H$_{2}$ molecules from the 
metal-nanotube complex, Na atom should remain in its equilibrium state. It may further be noted that the Na 
atom can transit from site to site easily on the carbon nanotube which facilitates the ultimate formation of 
high coverage.
\begin{table}
\center
\begin{center}
\caption{The binding energy per H$_{2}$ molecules (in~eV) and hydrogen weight percentage of the different SWCNTs}
\begin{tabular}{|l|c|c|c|c|c|}
\hline
Systems &       \multicolumn{2}{c|}{H$_{2}$} &  \multicolumn{2}{c|}{Na} &       \%~wt~  \\
\hline
&       LDA     &       GGA     &       LDA     &       GGA     &       \\      \hline
(3,3)SWCNT-Na$_{6}$-(H$_{2}$)$_{36}$    &       -0.24   &       -0.11   &       -1.26   &       -0.93
&       11.28 \\
(4,2)SWCNT-Na$_{7}$-(H$_{2}$)$_{42}$    &       -0.23   &       -0.10   &       -1.48   &       -1.18
&       9.20    \\
(5,0)SWCNT-Na$_{5}$-(H$_{2}$)$_{30}$    &       -0.24   &       -0.13   &       -1.73   &       -1.34
&       9.22   \\
(5,5)SWCNT-Na$_{10}$-(H$_{2}$)$_{60}$       &       -0.21   &       -0.09   &       -1.05    &      -0.85
&       11.28 \\
(6,3)SWCNT-Na$_{11}$-(H$_{2}$)$_{66}$       &       -0.23   &       -0.11   &       -1.12    &      -0.94
&       9.50 \\
(8,0)SWCNT-Na$_{8}$-(H$_{2}$)$_{48}$        &       -0.24   &       -0.08   &       -1.22    &      -0.86
&       9.20      \\   \hline
\end{tabular}
\end{center}
\end{table}

We have so far discussed the interaction of H$_{2}$ with a single atom bonded to a nanotube, but clearly one 
imagines attaching more Na to a nanotube, thereby increasing the hydrogen storage capacity. Figure 12 shows 
the optimized structures of all geometries considered here. Table 3 shows the binding energies of Na, H$_{2}$ 
molecules and weight percentage. We have observed that the binding energy~-0.24~eV of (8,0)SWCNT-Na$_{8}$-(H$_{2}$)$_{48}$
is larger than -0.17~eV/H$_{2}$ in (H$_{2}$)$_{64}$-Li$_{8}$-C$_{64}$ system with gravimetric density of 
13.45~wt\%\cite{jpc-Liu09}. In fact, these configurations store H$_2$ molecules approximately in the range
of 9.2--11.28~wt~$\%$. However, these numbers are based on the assumption that SWCNT-Na-H$_{2}$ group will also 
release the hydrogen molecule without difficulty. This assumption is perhaps a good one, since the binding 
energy of H$_{2}$ in SWCNT-Na-H$_{2}$ is about 1/8 of the binding energy of SWCNT-Na.
\begin{figure}
\center
\includegraphics[scale=0.2]{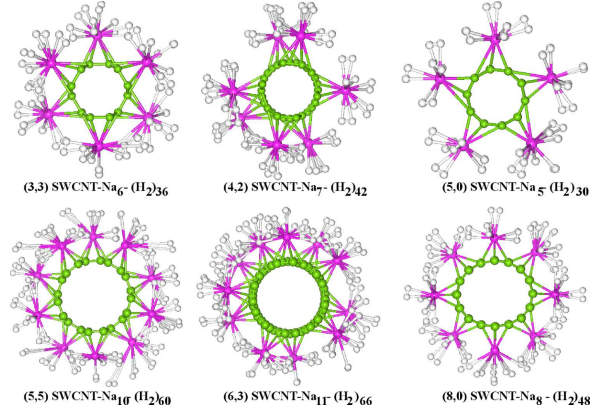}
\caption[small]{Optimized atomic structure of maximal number of adsorbed H$_{2}$ molecules on (3,3), (4,2), (5,0),
(5,5), (6,3) and (8,0) SWCNT-Na$_{n}$ systems.}
\end{figure}

Thus far, for the various states of the chemisorptions process, we have obtained the energy and geometry 
structure of atoms and molecules from first principle calculations at absolute temperature. A practical 
question thus arises as to how pressure and temperature effects will influence these stable and transition
states, and how the associated molecular vibrations vary over these energy states at given pressure and 
temperature. For an accurate understanding of the effects of pressure and temperature detailed thermodynamic 
study is required. However, such a study is beyond the scope of present work. Hence, we give a qualitative 
explanation of effect of pressure and temperature on the stability and storage capacity. It is noted that 
under the influence of high pressure, the internuclear separation decreases. The decrease in internuclear 
distances induces the confinement of electrons which result in localization and overlap of atomic orbitals, 
so the binding strength and storage capacity increases. This is supported by recent report on pressure 
dependence on hydrogen storage in Li-doped carbon nanotube\cite{catalysis-jung07}. The dependence of the 
stability of composite system (Na$_{n}$-SWCNT-(H$_{2}$)$_{n}$ on temperature and pressure is also explained 
with chemical potential which is given in terms of Gibbs free energy difference. The Gibbs free energy 
difference for the formation of stable adsorbates is given as follows;
\begin{equation}
\Delta{G}=E_{ads-swcnt}-E_{swcnt}-x\mu_{ads}
\end{equation}
where {\it{x}}, $\mu_{ads}$ are the hydrogen coverage and chemical potential of adsorbates 
respectively. The $\mu_{ads}$ depends on the temperature {\it{T}} and pressure {\it{P}} of the environment. 
The stable states are obtained by determining the minima under the same chemical potential. For an ideal gas,
$\mu_{ads}$ is given as 
\begin{equation}
\mu_{ads}=-k_{B}Tln\frac{(2\pi m)^\frac{3}{2}(kT)^\frac{5}{2}}{Ph^3}
\end{equation}
where $k_{B}$, m and {\it h} are Boltzmann constant, mass of adsorbate atoms, and Planck constant respectively. 
From above equation, we found that as the temperature approaches zero, $\mu_{ads}$~$\rightarrow$~0 and adsorbates 
coverage on the SWCNT's reaches a maximum. Here, we have $\mu_{ads}$$~<$~0 with increasing temperature, which 
results in decrease in the stability and storage capacity of adsorbates\cite{prb-yang06}.
\section{Conclusions}
In summary, using the state of the art first-principles total energy calculations at absolute temperature and 
pressure, we have shown that each Na atom adsorbed on SWCNT can bind up to six hydrogen molecules with an average 
binding energy of -0.26~eV/H$_{2}$. It is important to note that hydrogen molecules are found to interact with Na 
atom as a dihydrogen molecule, forming T-shaped complex and thus retaining their molecular identity irrespective 
of the type of SWCNTs. Our calculations show that the electric field produced by SWCNT-Na polarizes the H$_{2}$
molecules and the resulting induced dipole-dipole as well as the charge-dipole interaction, consequently, 
induces the binding of H$_{2}$ around SWCNT-Na. The PDOS of SWCNT-Na-H$_{2}$ shows that there is no bonding 
orbital between H$_{2}$ and Na, which also supports this argument. We have also investigated the effects
of nanotube chiral index (metallic and semiconducting) on Na and H$_{2}$ adsorption and found that semiconducting 
zigzag nanotubes, (8,0) and (5,0), show stronger binding energies of the Na atom, -2.1~eV and -2.6~eV 
respectively, as compared to metallic SWCNTs. We have shown that the Na adsorbates on SWCNTs are highly stable 
without the tendency for clustering at Na-Na distance of 5.9~\AA. Transition state calculations show that Na 
atom can easily move on the carbon nanotube and fill the nanotube exterior with metal and retain its stability during
desorption of H$_{2}$ molecules. At full coverage this yields a storage capacity of $\sim$ 9.2-11.28~wt~$\%$, 
which is higher than the value set for required gravimetric density of hydrogen storage. Our analysis shows that, 
although indeed Na-coated SWCNTs presents a potential material for the hydrogen storage, utter care should be 
taken to avoid Na atoms clustering on support material, to achieve higher hydrogen capacity.
\section{Acknowledgement}
This research is supported by Grant no. BRNS/BSC/37/38/3784 from the Department of Atomic Energy, Board of 
Research in Nuclear Sciences (DAE-BRNS), India. It is also supported in part by the Department of Science and 
Technology (DST) through computing resources provided by the High Performance Computing Facility 
at the Inter University Accelerator Centre (IUAC), New Delhi. SC acknowledges DST for providing financial assistance.
\bibliography{bibliography}

\begin{thebibliography}{10}

\bibitem{sch-01}
L.~Schlapbach and A.~Zuttel.
\newblock {\em Nature}, 414:353, 2001.

\bibitem{meng-07}
S.~Meng, E.~Kaxiras, and Z.~Zhang.
\newblock {\em Nano Letters.}, 7:663, 2007.

\bibitem{xiong-08}
Z.~T. Xiong, C.~K. Yong, G.~T. Wu, P.~Chen, W.~Shaw, A.~Karkamkar, A.~Autrey,
  M.~O. Jones, S.~R. Johnson, P.~P. Edwards, and W.~I.~F. David.
\newblock {\em Nat. Mater.}, 7:138, 2008.

\bibitem{nature-Dil97}
A.~C. Dillon, K.~M. Jones, T.~A. Bekkedahl, C.~H. Kiang, D.~S. Bethune, and
  M.~J. Heben.
\newblock {\em Nature}, 386:337, 1997.

\bibitem{heden-2000}
M. Heden, Minutes of meeting of the Hydrogen Technical Advisory panel 28-29
  Feb, 2000.

\bibitem{pavel-07}
Pavel~O. Krasnov, Feng Ding, Abhishek~K. Singh, and Boris~I. Yakobson.
\newblock {\em J. Phys. Chem. C Lett.}, 111:17977, 2007.

\bibitem{prl-Cir05}
T.~Yildirim and S.~Ciraci.
\newblock {\em Phys. Rev. Lett.}, 94:175501, 2005.

\bibitem{jpc-Liu09}
W.~Liu, Y.~H. Zhao, Y.~Li, Q.~Jiang, and E.~J. Lavernia.
\newblock {\em J. Phys. Chem.}, 113:2028, 2009.

\bibitem{science-Che99}
P.~Chen, X.~Wu, J.~Lin, and K.~L. Tan.
\newblock {\em Science}, 285:91, 1999.

\bibitem{jcp-Cab05}
I.~Cabria, M.~J. Lopez, and J.~A. Alonso.
\newblock {\em J. Chem. Phys.}, 123:204721, 2005.

\bibitem{jacs-sun06}
Q.~Sun, P.~Jena, Q.~Wang, and M.~Marquez.
\newblock {\em J. Am. Chem. Soc.}, 128:9741, 2006.

\bibitem{prl-yoon08}
M.~Yoon, S.~Yang, C.~Hicke, E.~Wang, D.~Geohegan, and Z.~Zhang.
\newblock {\em Phys. Rev. Lett.}, 100:206806, 2008.

\bibitem{prl-zhao05}
Y.~Zhao, Y.-H. Kim, A.~C. Dillon, M.~J. Heben, and S.~B. Zhang.
\newblock {\em Phys. Rev. Lett.}, 94:155504.

\bibitem{apl-shin06}
W.~H. Shin, S.~H. Yang, W.~A. Goddard, and J.~K. Kang.
\newblock {\em Appl. Phys. Lett.}, 88:053111, 2006.

\bibitem{carbon-Yan00}
M.~Kang.
\newblock {\em Carbon}, 38:623, 2000.

\bibitem{apl-Ataca08}
C.~Ataca, E.~Aktürk, S.~Ciraci, and H.~Ustunel.
\newblock {\em Appl. Phys. Lett.}, 93:043123, 2008.

\bibitem{nanolett-Gho08}
K.R.S. Chandrakumar and S.K.Ghosh.
\newblock {\em Nano Letters.}, 8:13--19, 2008.

\bibitem{chinphys-shi11}
Shi Guo-Sheng, Wang Zhi-Gang, Zhao Ji-Jun, Hu~Jun, and Fang Hai-Ping.
\newblock {\em Chinese Phys. B}, 20:068101, 2011.

\bibitem{langchak-95}
D.~V. Chakarov, L.~Osterlund, and B.~Kasemo.
\newblock {\em Langmuir}, 11:1201, 1995.

\bibitem{assglee-04}
M.~A. Gleeson, K.~Martensson, B.~Kasemo, and D.~Chakarov.
\newblock {\em Appl. Surf. Sci.}, 91:235, 2004.

\bibitem{rscsri-12}
K.~Srinivasu, Swapan~K. Ghosh, R.~Das, b~S.~Giri, and P.~K. Chattaraj.
\newblock {\em RSC Advances}, 2:2914--2922, 2012.

\bibitem{chemmater-Zheng08}
Shiyou Zheng, Fang Fang, Guangyou Zhou, Guorong Chen, Liuzhang Ouyang, Min Zhu,
  and Dalin Sun.
\newblock {\em Chem. Mater.}, 20:3954--3958, 2008.

\bibitem{jamchemsoc-cornelis08}
Cornelis~P. Balde, Bart~P.C. Hereijgers, Johannes~H. Bitter, and Krijn~P.
  de~Jong.
\newblock {\em J. Am. Chem. Soc.}, 130:6761--6765, 2008.

\bibitem{chandra-08}
K.R.S. Chandrakumar, K.~Srinivasu, and S.~K. Ghosh.
\newblock {\em J. Phys. Chem. C}, 112:15670, 2008.

\bibitem{prb-Per81}
J.~P. Perdew and A.~Zunger.
\newblock {\em Phys. Rev. B}, 23:5048, 1981.

\bibitem{prl-Cep80}
D.~M. Ceperley.
\newblock {\em Phys. Rev. Lett.}, 45:566, 1980.

\bibitem{prl-Per96}
J.~P. Perdew, K.~Burke, and M.~Emzerhof.
\newblock {\em Phys. Rev. Lett.}, 77:3865, 1996.

\bibitem{jcpbeck-88}
A.~D. Becke.
\newblock {\em J. Chem. Phys.}, 88:2547, 1988.

\bibitem{jcp-cab08}
I.~Cabria, M.~J. Lopez, and J.~A. Alonso.
\newblock {\em J. Chem. Phys.}, 128:144704, 2008.

\bibitem{jpcb-oka01}
Y.~Okamoto and Y.~Miyamoto.
\newblock {\em J. Phys. Chem. B}, B 105:3470, 2001.

\bibitem{jpcb-zho06}
Z.~Zhou, J.~J. Zhao, Z.~F. Chen, X.~P. Gao, T.~Y. Yan, B.~Wen, and P.~V.~R.
  Schleyer.
\newblock {\em J. Phys. Chem. B}, 110:13363--13369, 2006.

\bibitem{prb-lug07}
A.~Lugo-Solis and I.~Vasiliev.
\newblock {\em Phys. Rev. B}, 76:235431, 2007.

\bibitem{prb-lee08}
B.~Partoens O.~Leenaerts and F.~M. Peeters.
\newblock {\em Phys. Rev. B}, 77:125416, 2008.

\bibitem{prb-tad01}
S.~Furuya K.~Tada and K.~Watanabe.
\newblock {\em Phys. Rev. B}, 63:155405, 2001.

\bibitem{jmat-sah08}
U.~Sahaym and M.~G. Norton.
\newblock {\em J. Mater. Sci.}, 43:5395, 2008.

\bibitem{prb-gir02}
L.~A. Girifacol and M.~Hodak.
\newblock {\em Phys. Rev. B}, 65:125404, 2002.

\bibitem{zfk-Cla05}
S.~J. Clark, M.~D. Segall, C.~J. Pickard, P.~J. Hasnip, M.~J. Probert,
  K.~Refson, and M.~C. Payne.
\newblock {\em Zeitschrift fuer Kristallographie.}, 220(5-6):567--570, 2005.

\bibitem{jcp-li03}
J.~Li and S.~Yip.
\newblock {\em J. Chem. Phys.}, 119:2376, 2003.

\bibitem{catalysis-jung07}
Jung~Hyun Cho and Chong~Rae Park.
\newblock {\em Catalysis Today.}, 120:407, 2007.

\bibitem{prb-yang06}
Xiaobao Yang and Jun Ni.
\newblock {\em Phys. Rev. B}, 74:195437, 2006.

\end{thebibliography}
\bibliographystyle{unsrt}
\end{document}